\documentclass[11pt]{article}
\input{psfig}

\newcommand{\ket}[1]{\left | #1 \right \rangle}
\begin{document}

\begin{center}
{\small Appearing in {\em Proc. Roy. Soc. (Lond) A}, (2000)}\\[4mm]
{\Large\bf Counterfactual Computation\\ }
\bigskip
{\normalsize Graeme Mitchison$^\dagger$ and Richard Jozsa$^\S$ }\\
\bigskip
{\small\it $^\dagger$MRC Laboratory of Molecular Biology,Hills Road,
Cambridge CB2 2QH, UK. \\[2mm] $^\S$Department of Computer Science,
University of Bristol,\\ Merchant Venturers Building, Bristol BS8 1UB
U.K.}
\\[4mm]

\end{center}

\begin{abstract}
Suppose that we are given a quantum computer programmed ready to
perform a computation if it is switched on. Counterfactual computation
is a process by which the result of the computation may be learnt
without actually running the computer. Such processes are possible
within quantum physics and to achieve this effect, a computer
embodying the possibility of running the computation must be
available, even though the computation is, in fact, not run. We study
the possibilities and limitations of protocols for the counterfactual
computation of decision problems (where the result $r$ is either 0 or
1). If $p_r$ denotes the probability of learning the result $r$ ``for
free'' in a protocol then one might hope to design a protocol which
simultaneously has large $p_0$ and $p_1$. However we prove that $p_0 +
p_1 \leq 1$ in any protocol of this type and derive further
constraints on $p_0$ and $p_1$ in terms of $N$, the number of times
that the computer is not run. In particular we show that any protocol
with $p_0 + p_1= 1-\epsilon$ must have $N$ tending to infinity as
$\epsilon$ tends to 0. We show that ``interaction-free'' measurements
can be regarded as counterfactual computations, and our results then
imply that $N$ must be large if the probability of interaction is to
be close to zero. Finally, we consider some ways in which our
formulation of counterfactual computation can be generalised.
\end{abstract}

\begin{center}{\small\bf Keywords: quantum computation,measurement,
counterfactual,interaction-free} \end{center}

\section{Introduction}

There is a set of remarkable phenomena that seem to be special to
quantum mechanics. Their common theme is {\em counterfactuality}: the
fact that an event might have happened enables one to obtain some
information about that event, even though it did not actually take
place. Examples of such counterfactual phenomena include the
Elitzur-Vaidman bomb testing problem (Elitzur \& Vaidman 1993, Penrose
1994) and the use of so-called interaction-free measurements (Kwiat
{\em et al.} 1995, Renninger 1960, Dicke 1981, Geszti 1998) to determine the
presence or absence of an object by means of a test particle, even
though no ``interaction'' may have occurred between the object and the
particle. With some protocols, the probability of an interaction
occurring can be made arbitrarily small. By carrying out a raster-scan
of such measurements, it is possible to form an image of an object in
an interaction-free manner (White {\em et al.} 1998).

One can extend these ideas and imagine the object being replaced
by a quantum computer (Jozsa 1999). In that case, it turns out
that one can determine the outcome of a computation without the
computer ever being switched on. This is {\em counterfactual
computation}.  In the quantum formalism the computer may be in a
superposition of being on and off, and we will need to clarify the
sense in which the computer is ``not switched on''. This is the
content of our precise definition of counterfactual computation
given below. There are many possible protocols for counterfactual
computation which give various probabilities of gaining
information `for free'. The aim of this paper is to show that
there are limits on these probabilities which hold for all
possible protocols; there are also limits for protocols depending
on how often the computer is given the opportunity to run (in a
sense to be made precise). As we shall see, counterfactual
computation includes interaction-free measurement as a special
case, and our limits then translate into statements about the
probability of interaction.

It is well known that quantum physics has a profound bearing on issues
of computation as described for example in (Bernstein \& Vazirani
1993, Deutsch 1985, Deutsch \& Jozsa 1992, Ekert \& Jozsa 1996, Grover
1996, Simon 1994, Shor 1994). Much of that work is directed towards
questions of computational complexity, in particular devising quantum
algorithms that exhibit an exponential speedup over any known
classical algorithm for the computational task. Counterfactual
computation is a novel quantum computational effect of an entirely
different sort. It does not involve a speedup but highlights in a
particularly poignant way, some of the interpretational enigmas of
quantum mechanics and its theory of measurement. It remains to be seen
whether it can be exploited for real practical benefit.

\section{Defining counterfactual computation}

We begin by formalising the idea of counterfactual (henceforth
abbreviated to CF) computation. Consider a quantum computer with an
`on-off' switch programmed ready to solve a decision problem when
the switch is turned to `on'. This computer has an output register
that represents the binary result of the computation. The switch
and the output register will be denoted by a pair of qubits ${\cal
S}\otimes {\cal O}$, with $\cal S$ the switch qubit and $\cal O$
the output register. The switch states `off' and `on' are labelled
respectively as $\ket{0}$ and $\ket{1}$.

The computer will generally require extra storage qubits for its
programming and extra working qubits for its computational
processing. If $\cal R$ denotes the state space of these extra
qubits then the total state space of the computer is ${\cal
S}\otimes {\cal O}\otimes {\cal R}$. As is customary in idealised
quantum computation, we will assume that the computer operates in a
fully reversible manner and that the initial and final states of
$\cal R$ are equal. Also the binary result $r$ of the computation
is deposited in the output register qubit as an addition modulo 2.
Thus the computational process may be written
\begin{equation} \label{comp} \begin{array}{ll}
\ket{0}\ket{j}\ket{R} \rightarrow \ket{0}\ket{j}\ket{R} &
\mbox{(computer `off')}  \\
\ket{1}\ket{j}\ket{R} \rightarrow \ket{1}\ket{j\oplus r}\ket{R} &
\mbox{(computer `on')}
\end{array}
\end{equation}
and it is completed in some finite known time $T$.

If the computer is switched on, after a time $T$ it will have
carried out one of two unitary operations $U_0$ or $U_1$ on ${\cal
S}\otimes {\cal O}$ corresponding to the two possible outputs $r=0$
or $r=1$. Here $U_0$ is the identity on the switch and register
qubits and $U_1$ is the C-NOT operation on these two qubits (cf eq.
(\ref{comp})). Thus $U_0$ takes $|10\rangle$ into $|10\rangle$ and
$U_1$ takes $|10\rangle$ into $|11\rangle$ but if the computer is
not switched on then the evolution is the same for both possible
values of $r$ (in fact, being then $U_0$).

We will assume that the computer is given as a device with the
switch and output qubits ${\cal S}\otimes {\cal O}$ initially set
to $\ket{0}\ket{0}$. If the switch is set to $\ket{1}$ the computer
will effect either the transformation $U_0$ or $U_1$ and we wish to
determine which it is, without switching it on. We will assume that
we are unable to access the qubits in $\cal R$.

We begin with a row of qubits (as long as required) all initialised in
a standard state $\ket{0}$. A {\em protocol} consists of a sequence of
steps where each step is one of the following:\\ (a) A unitary
operation (not involving the computer) on a finite number of specified
qubits.\\ (b) A measurement on a finite number of specified
qubits.\\(c) An ``insertion of the computer'', where the state of two
selected qubits is swapped into the registers ${\cal S}\otimes {\cal
O}$ of the computer, a time $T$ is allowed to elapse and finally the
states are swapped back out into the two selected qubits.\\[1mm] The
steps of the protocol are implemented by selecting the designated
qubits from the row, applying the operation and then returning the
qubits to their positions in the row. The whole protocol may be
thought of as a network of operations with the computer being inserted
at various places via (c). The same $U_r$ is used throughout the
protocol. The outcomes of the measurements in (b) are the source of
our information about the computation and its result.

To understand how a protocol works and to provide a definition of
counterfactual computation, we introduce a notation which records
whether the computer was on or off on the various occasions when
it was inserted. At the beginning of any such step the entire
state space can be partitioned into two orthogonal subspaces, the
`off' and `on' subspaces, corresponding to the switch states
$|0\rangle$ and $|1\rangle$, respectively, and we will separate
the total state into its two superposition components in these
subspaces.

Imagine that, after each insertion of the computer, we carried out a
measurement that projects into these subspaces, with outcomes which we
write compactly as $f$ (for of{\bf f}) or $n$ (for o{\bf n}). Then
each possible list of $f/n$ outcomes, together with outcomes of the
measurements in the protocol, defines what we shall call a {\em
history}. One can depict this imaginary protocol as a branching
structure. At each node we give the {\em un-normalised} state vector,
the result of the projections occurring at each measurement step of
the protocol or at the imaginary measurement following an insertion of
the computer. If $v_n$ denotes the un-normalised state vector at the
final node of the history $h$, the probability of $h$ is
$|v_h|^2$. The initial node is assumed to correspond to the specified
initial state of the row of qubits. Note that there will be a
different branching structure depending on whether $U_0$ or $U_1$ is
being used.

In the real protocol, we do not necessarily carry out the $f/n$
measurement (though of course a protocol {\em could} include such a
measurement). Thus in contrast to actual measurement steps of type
(b), the unnormalised state vectors at the $f/n$ branchings still
retain the information of relative phases between them i.e. the
branching gives two coherent superposition components of the state
vector. Then we can utilise the branching structure of the imaginary
protocol to compute probabilities for the real protocol as
follows. Let $m$ denote a sequence of measurement outcomes, one for
each measurement step of type (b) in the protocol. There will be
various histories $h$ that include those outcomes (we write $m \subset
h$). For instance, in Example 1 below, with $U_0$ the histories
$f0f00$ and $n0f00$ both include the measurement outcomes
$m=0,00$. (Here the first measurement in the protocol is of one qubit,
giving the `0', whereas the second measurement is of two qubits,
giving `00'; see Figure 1). The probability of $m$ is just $|\sum_{m
\subset h} v_h|^2$, the sum being taken over all histories including
$m$. The contributions from these histories parameterised by $f/n$
sequences, are added coherently as the $f/n$ branchings do not
correspond to actual measurements. (If an $f/n$ measurement is
actually performed, its designated outcome will be listed in $m$
instead).

Note that our use of the term `history' differs from the more
conventional use (e.g. in the so-called consistent histories
approach to quantum mechanics). In the latter, every node is
viewed as a measurement and different paths (histories) are always
added incoherently (i.e. as a sum of squares rather than a square
of a sum) in computing the probability of a sequence $m$ of
specified outcomes of a subset of all the measurements performed.

We are now ready to define a CF computation. Let $m$ denote a
sequence of measurement outcomes. We say that $m$ is a CF outcome
of type $r$, where $r=0$ or $1$, if a history $h$ including $m$
satisfies $v_h \ne 0$ with $U_r$ only if $h$ is all-off, and if
the probability of $m$ with $U_{1-r}$ is zero. This means that if
we observe the measurement sequence $m$ then we can infer that the
computation result is certainly $r$ even though the computer has
not run at any stage of the protocol (since $m$ can only occur via
an all-off history). More formally we have:

{\em Definition: } $m$ is a CF outcome of type $r$, where $r=0$ or
1, if\\
1) With $U_r$, $v_h=0$ for all $h$ with $m \subset h$, except for that
$h$ that has only $f$'s (in addition to $m$).\\
2) With $U_{1-r}$, $\sum_{h, m \subset h} v_h=0$, the sum being over
all $h$ that include the measurement outcomes $m$; i.e.  with
$U_{1-r}$, $m$ is seen with probability zero.

Let us pause to elaborate on the sense in which ``the computer has
not been run at any stage'' if the CF outcome sequence $m$ is
observed. The reader may object that the computer has certainly
been in a superposition of `off' and `on' and in this sense, has
actually been run! But the sequence $m$ can arise only via all-off
histories so if it is seen then we have been confined to a part of
the total quantum state in which the computer is never run. For
example, in the language of the many worlds interpretation, we
will have evolved in a world in which the computer was never run
yet in that world we learn the result of the computation. The fact
that the computer may have run in another world is of no
consequence for {\em us}. The validity of this point is especially
emphasised in the original Elitzur-Vaidman bomb testing problem
(Elitzur \& Vaidman 1993) (which we will later see as a special case
of our CF formalism): there is always a world in which a good bomb
will explode but (with a suitable measurement outcome) we are
confined to another world in which the bomb is left {\it
unexploded} and yet we have the knowledge that it is, in fact, a
{\it good} bomb, available for future explosive applications.

This issue is at the heart of the so-called measurement problem of
quantum theory. Consider the proverbial Schr\"{o}dinger cat in
state $\frac{1}{\sqrt{2}}(\ket{\rm alive} +\ket{\rm dead})$. What
happens upon a measurement of alive versus dead? Suppose that the
outcome is seen to be ``alive''. Then in the conventional
``collapse of wavefunction'' formalism, the state of the cat
changes discontinuously into $\ket{\rm alive}$ and the $\ket{\rm
dead}$ component ceases to have any further physical reality or
existence or any further consequence. In a many-worlds or
decoherence view (Zurek 1991) of the measurement process both
components $\ket{\rm alive}$ and $\ket{\rm dead}$ persist and may
be thought of as existing in separate ``parallel'' worlds. The
measurer similarly passes into all these parallel worlds,
registering respectively all possible measurement outcomes. Since
$\ket{\rm alive}$ and $\ket{\rm dead}$ are macroscopically
distinct states they rapidly interact with many other subsystems
in the external universe, spreading orthogonality. This process is
the decoherence of the initial superposition which we symbolically
write as:
\begin{equation} \label{decoh} \begin{array}{lll}
\ket{\rm dead} \ket{0} \ket{0} \ldots \ket{0} & \rightarrow &
\ket{\rm dead} \ket{0} \ket{0} \ldots \ket{0} \\ \ket{\rm alive}
\ket{0} \ket{0} \ldots \ket{0} & \rightarrow & \ket{\rm alive}
\ket{1} \ket{1} \ldots \ket{1}
\end{array} \end{equation}
Here $\ket{0}$ and $\ket{1}$ are orthogonal states and the (very
long) string of kets after the leftmost ket represent a large
number of subsystems in the universe. Hence the two worlds or
branches will never again interfere {\em in practice} --
re-interference would require an enormous correlated effect to
undo the widespread interaction in eq. (\ref{decoh}) and is
extremely improbable. Hence although the cat exists dead in the
other world, we in {\em this} world (having perceived the outcome
``alive'') will never have to bother about or suffer any
consequences of that other unfortunate outcome. In either
interpretation we would be happy that the cat continues to exist
fully alive, despite that fact that it was in a strange
superposition in the past.

In a similar way in our definition of CF outcome, the desirable
option (like cat alive) is the computer being off and the
undesirable option (like cat dead) is the computer being on.
Although we pass through a superposition of these states, the
observation of a CF outcome means either that we have collapsed
the state to an entirely ``off'' part of the total state and the
``on'' part ceases to have any further physical reality or
consequences, or that we are in a world in which the computer has
not run, yet in both cases we also have the result of the
computation.

We note also that neither of the interpretations of the
measurement problem described above is really satisfactory: in the
first, no mechanism is given for the discontinuous ``collapse'' of
the state and in the second, no explanation is given of why we
actually perceive just {\em one} of the parallel worlds upon
making a measurement -- it is difficult to present tangible
physical evidence for the existence of the other worlds.\\

\noindent {\em Example 1}

\begin{figure}
\centerline{\psfig{figure=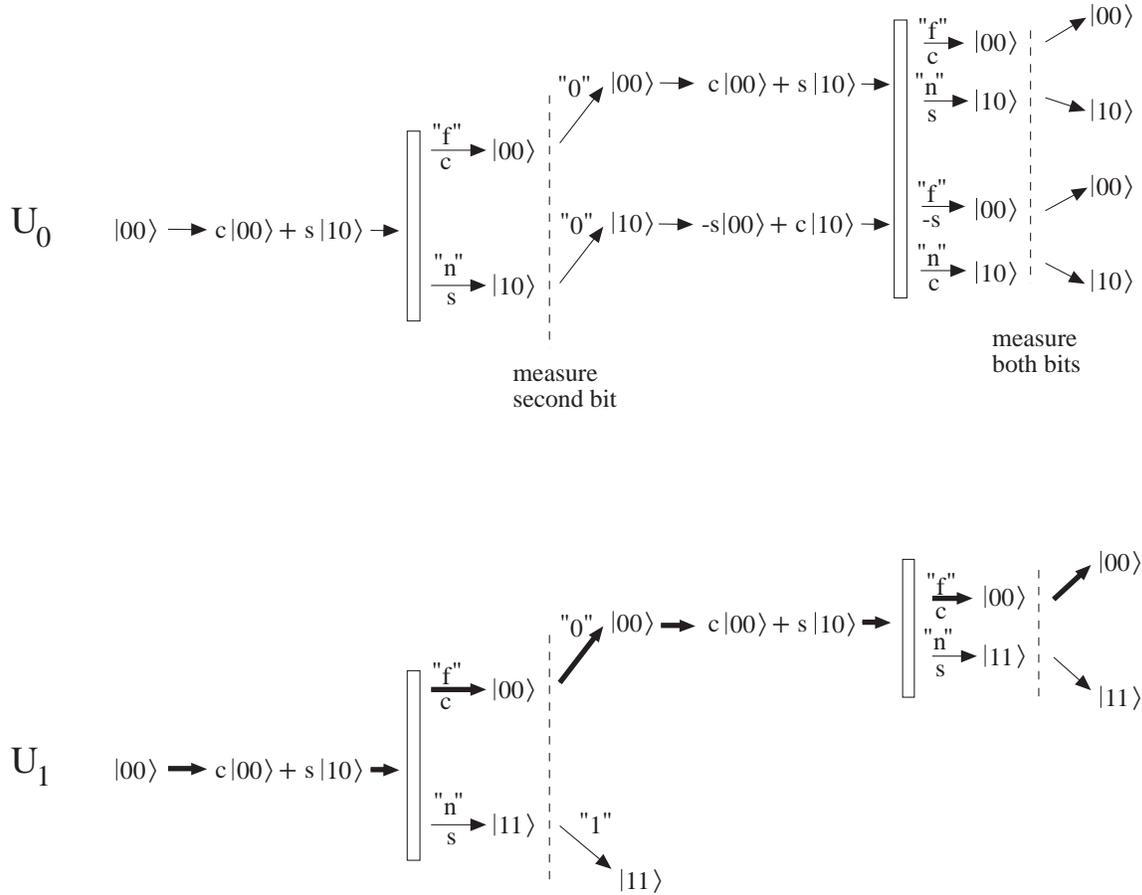,width=1.0\textwidth}} \caption{The
protocol for Example 1, for the case $N=2$, shown as a branching
structure, as described in the text. The thin boxes denote insertions
of the computer and the dotted vertical lines measurements. Here
$c=\cos \theta$ and $s=\sin \theta $. Arrows are labelled by $f$ or
$n$, and by measurement outcomes (these are all shown in inverted
commas). The sequence of labels defines the history associated to a
path. For instance, the uppermost path with $U_1$, marked with heavy
arrows, is the history $f0f00$. At each node the un-normalised state
vector is shown. Under $U_1$, this is the only history including the
measurement outcomes $m=0 \mbox{ and }00$, and it is an all-$f$
history, so condition 1) for a CF outcome of type 1 is satisfied.
Under $U_0$ there are two histories including $m$, namely $f0f00$ and
$n0f00$, and the sum of their final un-normalised state vectors is
$(c^2-s^2)\ket{00}$. If we chose $\theta=\pi/4$ so $c^2-s^2=0$, the
probability of $m$ under $U_0$ is zero, so condition 2) for a CF
outcome is satisfied. \label{fig:example1}}
\end{figure}

The following example is given in (Jozsa 1999) but we re-express
it here in terms of our formalism. The only variables that concern
us are the switch and output register qubits. Starting with
$|00\rangle$, the state is rotated to $\cos\theta|00\rangle+
\sin\theta|10\rangle$, where $\theta=\frac{\pi}{2N}$ for some
integer $N$. A time $T$ is then allowed for the computer to run,
if it is going to. This gives the state $\cos\theta|00\rangle+
\sin\theta|1r\rangle$, assuming $U_r$ is used ($r=0$, $1$). The
second qubit is now measured. If $r=1$, the measurement either
gives $|00\rangle$ with probability $\cos^2\theta$, or
$|11\rangle$ with probability $\sin^2\theta$. If the latter
occurs, we know that $r=1$ and the computer has run; the protocol
therefore halts. If $r=0$ the measurement yields
$\cos\theta|00\rangle+ \sin\theta|10\rangle$, and we repeat the
preceding steps, rotating by $\theta$, allowing the machine to
run, then measuring. The branching structures for $U_0$ and $U_1$
are shown in Figure \ref{fig:example1}.

After $N$ repeats, if $r=0$ the state will have rotated to
$|10\rangle$ with certainty. If $r=1$, the state will be
$|00\rangle$ with probability $\cos(\pi/2N)^{2N}$. In this case
the computer has not run, yet we know that $r=1$. This is
therefore an example of a CF computation. (In Figure
\ref{fig:example1}, the history that gives the CF result is shown
by bold arrows.) By making $N$ large enough, $\cos(\pi/2N)^{2N}$
can be brought arbitrarily close to 1. Thus if $r=1$ we obtain the
CF result with a probability approaching 1.\\

\noindent{\em Example 2}

\begin{figure}
\centerline{\psfig{figure=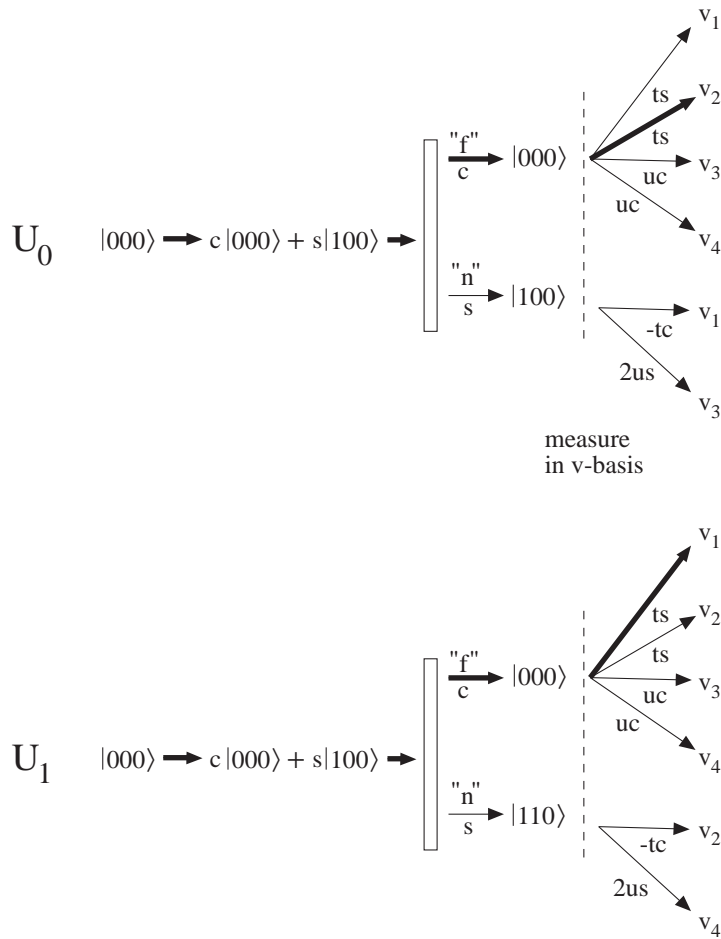,width=0.7\textwidth}} \caption{The
protocol for Example 2. Here the first two qubits are, as usual, the
switch and output register, and the third qubit corresponds to an
ancilla. The only vectors in the $x$-basis relative which have
non-zero probability of being final outcomes are those shown in the
figure. The bold arrows show the histories that give CF outcomes;
there is one of type 0 and one of type 1. \label{fig:example2}}
\end{figure}

We describe next a protocol where both types of CF outcome occur
(see Figure \ref{fig:example2}). In addition to the switch and
output qubits, we require a third, ancillary, qubit. To start
with, the initial state $|000\rangle$ is rotated to $c|000\rangle
+s|100\rangle$, where $c=\cos\theta$, $s=\sin\theta$. Next we
insert the computer on the first two qubits and finally measure
all qubits in the following basis:\\

{\noindent}$|x_1\rangle=t(s|000\rangle -c|100\rangle + s|001\rangle)$,\\
$|x_2\rangle=t(s|000\rangle - c|110\rangle - s|001\rangle)$, \\
$|x_3\rangle=u(c|000\rangle +2s|100\rangle + c|001\rangle)$, \\
$|x_4\rangle=u(c|000\rangle +2s|110\rangle - c|001\rangle)$, \\
$|x_5\rangle=|010\rangle$, $|x_6\rangle=|101\rangle$, \\
$|x_7\rangle=|011\rangle$, $|x_8\rangle=|111\rangle$,\\

where $t=1/\sqrt{1+s^2}$, $u=1/\sqrt{2+2s^2}$. There are two CF
outcomes, $\ket{x_1}$ with $r=1$ and $\ket{x_2}$ with $r=0$. They
satisfy our two conditions, namely: 1) the history leading to
$\ket{x_1}$ under $U_1$ has only `f's, as does that leading to
$\ket{x_2}$ under $U_0$, and 2) $\ket{x_1}$ has probability zero under
$U_0$ as does $\ket{x_2}$ under $U_1$. The probabilities of these CF
outcomes are both $(cst)^2$. This is maximised by $c^2=2-\sqrt2$ which
gives $(cst)^2=0.172$, or $p_0+p_1=0.344$.

\section{Limits on counterfactual computation}

Let us write $p_0$, $p_1$ for the probabilities of getting a CF
outcome of type 0 or type 1, respectively in any given protocol.
Example 1 shows that there is a protocol for which $p_1$
approaches 1, but in this case $p_0=0$. Could one devise a
protocol that allowed both $p_0$ and $p_1$ to approach 1? Less
ambitiously, can one devise a protocol giving $p_0+p_1>1$? These
questions were posed in Jozsa (1999), and we answer them here.

First note that, since the subdivision into on-off subspaces
corresponds to an orthogonal decomposition of the entire state
space (and the same is true of each measurement step of type (b)),
we have, for each $U_r$
\begin{equation}
\sum_h |v_h|^2=1.
\label{eq:unity}
\end{equation}
In Figure 1, for instance, under $U_0$ we have $\sum_h
|v_h|^2=c^4+c^2s^2+s^4+c^2s^2=1$, for any $\theta$. This is true
despite the fact that the final states of the histories may not be
orthogonal for the original protocol e.g. $c^2 \ket{00}$ and $-s^2
\ket{00}$ for $U_0$ in figure 1; we are using the fact that each
step in the histories is either a measurement in the protocol, or
can be regarded as one (as in our imaginary protocol).

This basic fact may be alternatively expressed as follows. We imagine
a supply of extra qubits, as many of them as there are insertions of
the computer in our protocol. If we carry out a C-NOT operation
between the switch qubit and the $n$-th extra qubit at the $n$-th
insertion, the resultant states at each branching of the imaginary
protocol will be mutually orthogonal in the extended space. Hence the
branching tree corresponds to a probabilistic process, and
eq. (\ref{eq:unity}) expresses the certainty of following {\it
some} path through the tree.

Now we know from condition 1) above that the all-off history $h$ alone
determines the probability $|v_h|^2$ of the CF outcome. But, for an
all-off history $h$, $v_h$ must be the same whether
$U_0$ or $U_1$ is used, {\em since the computer is never switched on!}
Thus, in forming the sum $p_0+p_1$, i.e. in collecting terms $|v_h|^2$
for all possible CF outcomes, one can assume that all of these terms
come from the list of $v_h$'s for one of the $U_r$, say $U_0$. Note
that condition 1) implies that there is no possible overlap: one can
only get a CF outcome with one of the $U_r$ for a given $m$.

Thus $p_0+p_1=\sum_{h \in \cal F} |v_h|^2$, where the sum is over the set
$\cal F$ of all-off terms from CF outcomes. These terms can be assumed to
be derived using $U_0$, so equation (\ref{eq:unity}) implies:

\noindent {\bf Theorem 3.1.}  $p_0+p_1 \le 1$.

This answers the conjecture in Jozsa (1999). We now look at some
subsidiary questions.

In Example 1, one can only approach the bound of 1 closely by having a
large number, $N$, of insertions of the computer. Could there be
protocols which reach the bound of 1, or come close, with only a few
insertions?  The following argument shows that this is not possible.

Consider first the situation shown in Table \ref{tab:cf1}. There is
only one CF outcome, and we ask how many insertions of the
computer we need to get the probability of this result, $p_1$, to
satisfy $p_1=1-\epsilon$.

\begin{table}
\caption{One CF outcome, $M_1$.
\label{tab:cf1}}
\begin{tabular}{|c|c|c|}
\hline
& $U_0$& $U_1$\\
\hline
$M_1$ & 0 & off only\\
\hline
$M_2$ & x & x \\
\hline
.. & .. & .. \\
\hline
\end{tabular}
\end{table}

Let $h(i)$ denote a history consistent with measurement outcome
sequence $M_i$. Under $U_0$ we have:
\[
\sum_{h(1)} |v_{h(1)}|^2=1-\sum_{h(i), i\ge2}|v_{h(i)}|^2 \le 1.
\]
Let $F(1)$ denote the all-$f$ history consistent with $M_1$. Then
\[
p_1=|v_{F(1)}|^2 \le 1-\sum_{h(1) \ne F(1)} |v_{h(1)}|^2,
\]
so,
\begin{equation}
\sum_{h(1) \ne F(1)}|v_{h(1)}|^2 \le \epsilon.
\label{eq:eps}
\end{equation}
On the other hand the total un-normalised state vector (under $U_0$
still) for $M_1$ is zero, so
\begin{equation}
\sum_{h(1)} v_{h(1)}=0.
\label{eq:zerosum}
\end{equation}
Thus
\begin{equation}
\sum_{h(1) \ne F(1)} v_{h(1)}=-v_{F(1)},
\label{eq:f1}
\end{equation}
and hence
\begin{equation}
\left|\sum_{h(1) \ne F(1)}v_{h(1)}\right|^2=|v_{F(1)}|^2=1-\epsilon.
\label{eq:moref1}
\end{equation}
Now, for any vectors $x_i$, the minimum of $\sum_{i=1}^K |x_i|^2$
given $|\sum_{i=1}^K x_i|^2=S$ is $S/K$. There are $2^N$
histories consistent with $M_1$ for $N$ computer insertions, so
\[
\sum_{h(1) \ne F(1)} |v_{h(1)}|^2 \ge (1- \epsilon)2^{-N}.
\]
Using equation (\ref{eq:eps}), $(1-\epsilon)2^{-N} \le \epsilon$, or
\begin{equation}
N \ge \log_2 \left(\frac{1-\epsilon}{\epsilon} \right).
\label{eq:Nbound}
\end{equation}

\begin{table}
\caption{Both CF outcomes: $r=1$ ($M_1$)
and $r=0$ ($M_2$).
\label{tab:cf2}}
\begin{tabular}{|c|c|c|}
\hline
& $U_0$& $U_1$\\
\hline
$M_1$ & 0 & off only\\
\hline
$M_2$ & off only & 0 \\
\hline
$M_3$ & x & x \\
\hline
.. & .. & .. \\
\hline
\end{tabular}
\end{table}

Consider next the situation shown in Table \ref{tab:cf2}. Here both
types of counterfactual result occur, and we assume
$p_0+p_1=1-\epsilon$. Under $U_0$ we have
\[
|v_{F(1)}|^2+|v_{F(2)}|^2+\sum_{h(1) \ne F(1)} |v_{h(1)}|^2 \le 1.
\]
Since $p_0=|v_{F(1)}|^2$ and $p_1=|v_{F(2)}|^2$, we obtain (\ref{eq:eps})
as before. Furthermore, equations (\ref{eq:zerosum}) and (\ref{eq:f1})
hold, but now the equivalent of (\ref{eq:moref1}) is
\begin{equation}
\left|\sum_{h(1) \ne F(1)}v_{h(1)}\right|^2=|v_{F(1)}|^2=p_1.
\end{equation}
We therefore obtain, in place of (\ref{eq:Nbound}), $N
\ge\log(p_1/\epsilon)$. On the other hand, under $U_1$ we have $N \ge
\log(p_0/\epsilon)$. Since $\max(p_0,p_1) \ge (1-\epsilon)/2$ we
conclude that
\begin{equation}
N \ge \log_2 \left(\frac{1-\epsilon}{2\epsilon} \right).
\label{eq:Nbound2}
\end{equation}
In Tables \ref{tab:cf1} and \ref{tab:cf2} only one CF outcome of
each type is shown. However, no real extra generality is gained by
allowing several CF outcomes of the same type. To see this, note
that in each measurement step of type (b), instead of collapsing
the state, we may leave it entangled with an auxiliary system
representing the measurement pointer. Thus each measurement step
becomes a unitary operation of type (a). Finally at the {\it end}
of the protocol we read all of the pointers together. This may be
viewed as a single measurement whose possible outcomes are the
possible measurement sequences of the original protocol. All CF
outcomes of one type can then be grouped into a single final
measurement outcome.

We summarise these results as follows:

\noindent {\bf Theorem 3.2.} {\em The number of insertions $N$ must
tend to infinity as $p_0+p_1$ tends to 1.}

It is possible that the inequalities
(\ref{eq:Nbound}) and (\ref{eq:Nbound2}) can be sharpened. This is
certainly true when $N=1$. Consider first the situation where only one
type of CF outcome occurs (Table \ref{tab:cf1}). The only possible
histories are $fi$ or $ni$, where $i$ denotes outcome $M_i$. With
$U_0$, (\ref{eq:unity}) gives
\[
\sum_i |v_{fi}|^2 + |v_{ni}|^2=1.
\]
Since $v_{f1}+v_{n1}=0$ we have
\begin{equation}
2|v_{f1}|^2=1-\sum_{i>=2} \left( |v_{fi}|^2 + |v_{ni}|^2 \right).
\label{eq:one}
\end{equation}
We can also add the un-normalised state vectors leading to $M_1$,
$M_2$, etc, and consider just the projections onto the subspaces of
the final measurement. This gives:
\begin{equation}
\sum_{i>=2} |v_{fi} + v_{ni}|^2=1,
\label{eq:two}
\end{equation}
where the sum goes from $i=2$ because $v_{f1}+v_{n1}=0$. From equations
(\ref{eq:one}) and (\ref{eq:two}) we get
\begin{equation}
p_1=|v_{f1}|^2=1/4-1/2 \sum_{i>=2} |v_{fi} - v_{ni}|^2 \le \frac{1}{4}.
\label{eq:onequarter}
\end{equation}
For $N=1$, (\ref{eq:Nbound}) implies $\epsilon \ge 1/3$, or $p_1 \le
2/3$, so (\ref{eq:onequarter}) is an improvement; in fact,
(\ref{eq:onequarter}) is best possible because Example 1 with $N=1$
gives $p_1=1/4$. (More precisely we would need an extra final rotation
step, omitted from figure \ref{fig:example1}, at the end of the
protocol before measuring both bits.)

Consider next the case where both types of CF outcome occur (Table
\ref{tab:cf2}). With $U_0$, (\ref{eq:unity}) gives
\[
\sum_i |v_{fi}|^2 + |v_{ni}|^2=1.
\]
and now $v_{f1}+v_{n1}=0$ and $v_{n2}=0$, so
\begin{equation}
2|v_{f1}|^2+|v_{f2}|^2=1-\sum_{i>=3} \left(|v_{fi}|^2 + |v_{ni}|^2 \right).
\label{eq:povmone}
\end{equation}
Again, we can add the un-normalised state vectors leading to $M_1$,
$M_2$, etc, which gives:
\begin{equation}
|v_{f2}|^2=1-\sum_{i>=3} |v_{fi} + v_{ni}|^2,
\label{eq:povmtwo}
\end{equation}
Equations (\ref{eq:povmone}) and (\ref{eq:povmtwo}) imply
\[
4|v_{f1}|^2+|v_{f2}|^2=1-\sum_{i>=3} |v_{fi}-v_{ni}|^2.
\]
With $U_1$ we get
\[
|v_{f1}|^2+4|v_{f2}|^2=1-\sum_{i>=3} |v_{fi}-\tilde v_{ni}|^2.
\]
the tilde in $\tilde v_{ni}$ distinguishing the state vector with
$U_1$ from that with $U_0$.  Adding the last two equations gives:
\[
5|v_{f1}|^2+5|v_{f2}|^2=2-(\mbox{positive terms}),
\]
so
\begin{equation}
p_0+p_1 \le 2/5.
\label{eq:twofifths}
\end{equation}
For $N=1$, (\ref{eq:Nbound2}) implies $\epsilon \ge 1/5$, or $p_0+p_1
\le 4/5$, so (\ref{eq:twofifths}) is an improvement. We do not know,
however, whether it is best possible. In Example 2,
$p_0+p_1=0.344$. Can we attain the bound of (\ref{eq:twofifths}) with
some other protocol?

\section{`Interaction-free'  measurements}

Our formalisation of counterfactual computation provides a general
framework that includes `interaction-free' measurements (IFM) (Elitzur
\& Vaidman 1993, White {\em et al.} 1998, Vaidman 1996) as a special
case which is characterised as follows. Suppose we have an apparatus
and an object. We interpret the switch qubit as defining
configurations in which the apparatus cannot interact with the object
(switch=$\ket{0}$), or can (switch=$\ket{1}$), and we interpret the
output register qubit as defining the state of the object, with
$\ket{0}$ signifying `no interaction' and $\ket{1}$ `interaction has
occurred'. The additional constraint that defines an IFM is that each
insertion of the computer, now viewed as a potential interaction
between apparatus and object, is followed by a measurement of the
output register. If this is $\ket{1}$, an interaction has occurred and
the protocol halts. In contrast, in a general CF protocol, the output
register may be set to $\ket{1}$ also by other unitary steps (not
involving the computer) and the $\ket{1}$ branches may be used in
subsequent interferences.

In an IFM protocol we interpret $U_0$ as the computation
corresponding to the object being absent, and $U_1$ to that when
it is present. Thus $p_1$ is the probability of an IFM occurring.
As for $p_0$, it can be interpreted as the probability of the
system being confined to the non-interactive configuration when
the object is absent. This is a rather artificial concept, and
theorem 3.1 does not translate into a very meaningful result. More
interesting is the implication of theorem 3.2:

\noindent {\bf Theorem 4.1.} {\em  The number of times $N$ an
interaction with an object does not occur must tend to infinity as
the probability of an IFM tends to 1.}

Our Example 1 can be interpreted as an IFM, since the output register
is measured after every insertion. Example 2 cannot be so interpreted,
however, since the measurement following the insertion leaves the
output register in a superposition of the states $\ket{0}$ and
$\ket{1}$.

\section{More general CF computation}

Our definition of CF computation may be generalised in various
natural ways such as the following. We may be given more than two
unitary operations $U_r$ on a product space ${\cal S}\otimes {\cal
O}$, whose parts may be larger than just qubits. Each operation
$U_r$ is required to have and ``on'' and ``off'' subspace defined
by an orthogonal partition of $\cal S$ and $U_r$ is the identity
on its ``off'' subspace. Given an unknown one of the $U_r$'s we
wish to determine $r$ while being confined to its ``off''
subspace.

We now describe an explicit example of this type of generalisation
which allows the switch space $\cal S$ to have an arbitrary number
of dimensions, and which also allows more than two operations
$U_r$, but $\cal O$ is still taken to be a qubit. We consider
later in more detail the case where there are only two $U_r$'s but
an arbitrary dimensional switch space, and prove an analogue of
theorem 3.2.

Our example is motivated by thinking of an interferometer in which
the incoming photon is split into an equal superposition of $K$
arms, numbered 0 to $K-1$, with absorbing objects blocking all the
arms except for arm 0 and arm $r$ ($r \ne 0$). The problem is to
identify the value of $r$ without an interaction occurring. Our
protocol below will also be an IFM in the sense of section 4 i.e.
the protocol measures the output register after each potential
interaction and halts if an interaction is seen to have occurred.
However this IFM problem differs from standard IFM, where no prior
assumption is made about the object being present or absent; here
we consider a special set of configurations of objects.

More formally we consider a $K$ dimensional switch with state space
$\cal S$ spanned by $\{ \ket{0}, \ldots , \ket{K-1} \}$. The output
register $\cal O$ is still a qubit and we define $K-1$ unitary
operators $U_r$ for $r= 1, \ldots , K-1$ on ${\cal S}\otimes {\cal O}$
by:\\

\noindent $U_r(\ket{ij})=\ket{ij}$, if $i=0, r$  (the `off'-subspace),\\
$U_r(\ket{ij})=\ket{i,1-j}$, if $i \ne 0, r$  (the `on'-subspace).\\

Thus $U_r$ corresponds to arms 0 and $r$ being unblocked. The
``off'' subspace for $U_r$ (on which $U_r$ is the identity) is
${\rm span}(\ket{0},\ket{r})\otimes {\cal O}$ while the ``on''
subspace (on which $U_r$ negates the output qubit value) is the
orthogonal complement. Given an unknown one of the $U_r$'s we wish
to identify $r$ while confining ourselves to its ``off'' subspace.
More formally we modify our original definition of CF outcome as
follows:

{\em Definition: } $m$ is a CF outcome of type $r$ if\\ 1) With
$U_r$, $v_h=0$ for all $h$ with $m \subset h$, except for that $h$
that has only $f$'s.
\\ 2) With $U_{s}$, $s \ne r$, $m$ is seen with probability zero.

Consider now the following protocol. Starting with $\ket{00}$, we
apply to the switch a unitary transformation $R$ given by
\[
R(\ket{00})=a\ket{00}+\sum_{i=1}^{K-1} b\ket{i0},
\]
and
\[
R(\ket{j0})=-b\ket{00}+\ket{j0}+\sum_{i=1}^{K-1} \frac{(a-1)}{(K-1)}\ket{i0}, j \ne 0.
\]
where $a^2+(K-1)b^2=1$. We now insert $U_r$, then measure the output
register. If it is $\ket{1}$ the protocol terminates; otherwise the
protocol continues by repeating the above. After $\pi/2b$ steps%
\footnote{More precisely, the number of steps that yields the
state closest to $\ket{r0}$ can be calculated for given $K$ and
$b$, and tends to $\pi/2b$ as $Kb$ tends to zero.}  the protocol
halts. We may show (cf the geometric interpretation of the process
below) that as $b$ decreases, the probability of interaction
occurring decreases. In a way that is similar to example 1, for
small enough $Kb$ the final state is, with high probability,
$\ket{r0}$. According to our definition above, this will then be a
CF outcome of type $r$.

The workings of the protocol may be understood in terms of a
geometrical interpretation of its operations. Consider first the
case $K=3$. Looking at the switch variable, $R$ is just a rotation
in the plane of $\ket{0}$ and $\ket{1}+\ket{2}$, i.e. about the
axis $\ket{1}-\ket{2}$ which is orthogonal to those two states. To
see this just note that $R$ is {\em some} rotation (as columns are
orthonormal and det$(R) =1$) and $R$ leaves $\ket{1}-\ket{2}$
fixed. The angle of rotation is given by $\cos \theta = \langle
v|Rv \rangle$ for any choice of $v$ in the rotation plane; taking
$v= \ket{0}$ gives $\cos \theta = a$. The protocol can then be
thought of as follows: $R$ repeatedly applied gradually pushes the
state $\ket{0}$ around to the equal superposition of all the other
basis states, except that after each incremental rotation $R$, the
state is projected into the 2D space of $\ket{0}$ and $\ket{r}$
(so long as the output register is always measured to be
$\ket{0}$). It is clear that $\ket{0}$ will be pushed around to
$\ket{r}$ and we get $p_r$ near $1$ for all $r$'s if $b$ is
sufficiently small, keeping the state always close to the rotation
plane. This 2D motion is not a simple rotation because of the
repeated projections (which do not commute with $R$). If the
output register is ever measured to be $\ket{1}$, the state has
been projected orthogonal to the rotation space and we abort the
process.

For general $K$ the operation $R$ on the switch variable is just the
same rotation in the 2D subspace of $\ket{0}$ and $\ket{1}+\ket{2} +
\ldots \ket{K-1}$, being the identity in the orthogonal
$(K-2)$-dimensional complement $C$. To see this note that $R$ leaves
invariant all $K-2$ vectors of the form $\ket{1}-\ket{r}$, $r=2,
\ldots , K-1$. These vectors are all orthogonal to both $\ket{0}$ and
$\ket{1}+\ket{2} + \ldots \ket{K-1}$ so they span $C$. The angle of
rotation is as given above and again we get a 2D motion in the plane
of $\ket{0}$ and $\ket{r}$.

Now consider $K=3$ again: As in our original definition of CF
computation, there are two operations, here $U_1$ and $U_2$, but
now the switch variable's dimension is three (i.e. the switch is a
``qutrit''). Both $p_1$ and $p_2$ have a natural interpretation,
(the probability of an IFM if an object is in arm 2 or 1,
respectively), so one might hope that theorem 3.1 would yield a
meaningful result. However, the theorem (stating that $p_1 + p_2
\leq 1$) evidently fails in this generalised setting since both
$p_1$ and $p_2$ tend to 1 for small enough $b$. The reason for
this failure is that the proof assumes that the two $U_r$'s have
the same `off' subspace. This was true of the original $U_r$'s but
is not true in the present case as the switch qutrit is $\ket{0}$
or $\ket{1}$ in the `off' space of $U_1$, and is $\ket{0}$ or
$\ket{2}$ in the `off' space of $U_2$. We therefore cannot hope to
prove an equivalent of theorem 3.1. However, we show next that a
version of theorem 3.2 {\em does} hold, namely that the number of
steps must be large for $p_0+p_1$ to approach its upper bound.
This bound is 2 in the present example rather than 1.

This new theorem holds not only for IFMs, but for any CF
computations, given certain restrictions on the `on'- and
`off'-subspaces of the $U_r$'s. We label the two $U_r$'s as $U_0$
and $U_1$ again, but now allow any finite dimensional switch space
${\cal S}$. We assume that $\cal S$ is partitioned into orthogonal
`on'- and `off'-subspaces for each of $U_0$ and $U_1$, and we also
assume that these decompositions are compatible in the sense that
the subspaces \\

\noindent ${\cal S}_a$= [$U_0$ `off' \& $U_1$ `on'], \\
${\cal S}_b$= [$U_0$ `on' \& $U_1$ `off'], \\
${\cal S}_f$= [$U_0$ `off' \& $U_1$ `off'], \\
${\cal S}_n$= [$U_0$ `on' \& $U_1$ `on'],
\\

\noindent span the whole of ${\cal S}$. We also assume that $U_0$ is the
identity on ${\cal S}_a\otimes {\cal O}$ and ${\cal S}_f\otimes {\cal
O}$, and $U_1$ is the identity on ${\cal S}_b\otimes {\cal O}$ and
${\cal S}_f\otimes {\cal O}$. With these assumptions we have

\noindent {\bf Theorem 5.1.} {\em The number $N$ of insertions must
tend to infinity as $p_0+p_1$ tends to its maximum value.}

The proof is given in the Appendix. Note that the theorem applies
to the 3-armed interferometer as ${\cal S}_a=\{\ket{1}\}$, ${\cal
S}_b=\{\ket{2}\}$, ${\cal S}_f=\{\ket{0}\}$ and ${\cal
S}_n=\emptyset$, and these subspaces span {\cal S}. Here there are
two operations $U_r$, but it seems likely that theorem 5.1 can be
generalised to any number of operations satisfying appropriate
compatibility conditions on their `on' and `off' subspaces.

\section{Discussion}

Our starting point was the enticing notion of being able to run a
quantum computer `for free'. The quotation marks here were very
necessary, however, for it is not at all clear what if anything
comes for free. A CF protocol requires the computer to be present,
and due time must allowed for the machine not to run. If the
computer does not run, then we might expect to gain protection
against decoherence. However, this appears not to be the case
because our computer has to eschew all dissipative processes if
running it is to constitute a unitary operation (as is necessary
in a CF protocol to correctly generate the destructive
interferences which allow us to conclude that an outcome is CF for
a value of $r$), even though it might not actually be run. Perhaps
the computer owner could determine whether the machine had run and
levy a charge accordingly; we would then at least make a financial
gain from a CF computation. Suppose, for instance, at each step
where there was an `on'-`off' choice the owner arranged for the
switch state to be entangled with an extra qubit. At the end of
the run he could measure the extra qubits to see what path the
computation had taken. However, it is easy to see (in example 1
for instance) that this measurement generally destroys the
interference that gives rise to a CF outcome.

Thus the naive notion of gain from a general CF computation has to
be abandoned. It does not follow, however, that CF computation is
a meaningless abstraction. As we have seen, the framework includes
`interaction-free' measurement, where (despite the quotation
marks) the practical gain is real, since it includes such
potential benefits as X-ray images with reduced radiation damage
(Vaidman 1996). By making the definition of CF computation
sufficiently broad, we might hope to capture other types of
interesting counterfactuality. In the last section we took some
steps towards generalising our initial definition. We conclude by
suggesting two further steps in this direction. These are both
probabilisitic extensions of CF computation.

First, suppose the output register is extended to $K+1$ values, and
that running the computer takes the state $|10\rangle\otimes|x\rangle$
to $|1r\rangle\otimes|x\rangle$ under $U_r$, for $r=0, \ldots, K$,
where $|x\rangle=|0 \ldots 0\rangle$ in a $2^K$-dimensional ancillary
Hilbert space. By analogy with Example 2, one can rotate
$|00\rangle\otimes|x\rangle$ to
$(c|00\rangle+s|10\rangle)\otimes|x\rangle$, insert the computer, then
measure in a basis including
$|v_r\rangle=t(s|00\rangle-c|1r\rangle)\otimes|x\rangle+ts|00\rangle
\otimes |y_r\rangle$, where the $|y_r\rangle$ are vectors from the
barycentre of a $K$-simplex to its vertices satisfying $\langle
y_i|y_j \rangle=-1$, $|y_i|^2=K$ and $t=1/\sqrt{1+Ks^2}$. If we see
the outcome corresponding to $|v_s\rangle$, we know that the output
register did {\em not} take the value $s$, and also that the computer
did not run. Thus we have some counterfactual information, but do not
know precisely what value the output register took. However, given
prior probabilities for the $U_r$, we can compute posterior
probabilities for output $r$.

The other type of probabilistic extension is obtained by relaxing the
conditions (1) and (2) in the definition of CF computation. We may
define a measurement outcome sequence $m$ to be ``approximately CF
for $r$'' if it has the following properties. For some (small)
$\epsilon$:\\ ($1^\prime$) With $U_r$, $\sum |v_h |^2 <\epsilon$
where the sum is over all histories consistent with $m$ except for
the all-off history (the latter history having probability
significantly larger than $\epsilon$).\\ ($2^\prime$) With $U_{1-r}$,
$m$ is seen with probability less than $\epsilon$.\\ Hence if $m$ is
seen then with high probability the computational result is $r$ and
also with high probability, no computation has been done.\\[3mm]
{\Large\bf Appendix A.}\\[3mm] We give here a proof of theorem 5.1.

\noindent {\em Proof.}  The upper bound here is 2, so suppose
$p_0+p_1=2-\epsilon$, for small $\epsilon > 0$.

\noindent We decompose ${\cal S} \otimes {\cal O}$ into subspaces
${\cal S}_r \otimes {\cal O}$, where $r=a, b, f, n$, and label each
component before an insertion with the appropriate subscript. Just as
we previously wrote histories as chains of $f$'s and $n$'s (and
measurement outcomes, which we suppress), we can now write histories
as chains of $a$'s, $b$'s, $f$'s and $n$'s. We have
\begin{equation}
p_0 \le |\sum_{a,f} v_s|^2,
\label{eq:p0def}
\end{equation}
where $s$ ranges over all strings composed of $a$'s and $f$'s, and
$v_s$ is obtained under $U_0$. Similarly for $p_1$ we have,
\begin{equation}
p_1 \le |\sum_{b,f} v_s|^2,
\label{eq:p1def}
\end{equation}
with $b$'s and $f$'s, under $U_1$.

Now consider a protocol with $N$ insertions. After the first insertion
(\ref{eq:p0def}) and (\ref{eq:p1def}) imply

$p_0 \le |v_a + v_f|^2 = |v_a|^2 + |v_f|^2$,  (under $U_0$)\\
$p_1 \le |v_b + v_f|^2 = |v_b|^2 + |v_f|^2$,  (under $U_1$).

Also,
\[
|v_a|^2 +|v_b|^2 +|v_f|^2 \le 1,
\]
where the first term is derived under $U_0$, the second under $U_1$
and the third under either. This holds because the $U_r$ are all the
identity on the subspaces in question. These inequalities imply
$|v_f|^2 \ge p_0+p_1-1$ and $|v_a|^2 + |v_b|^2 \le 2-p_0-p_1$, which
implies $|v_a|+|v_b| \le \sqrt{4-2p_0-2p_1}$. Since
$p_0+p_1=2-\epsilon$, we get

$|v_f|^2 \ge 1-\epsilon$.\\
$|v_a|+|v_b| \le \sqrt{2\epsilon}$.

Note that the first inequality implies ${\cal S}_f$ is non-empty.

Now look at the situation after n insertions. From
(\ref{eq:p0def}) we have, under $U_0$:
\begin{equation}
p_0 \le |\sum_{k=0}^{k=n-2}\sum_s v_{f^kas^{n-k-1}} + v_{f^{n-1}a} +
v_{f^n}|^2.
\label{eq:p0full}
\end{equation}
Here $f^kas^{n-k-1}$ denotes a history beginning with $k$ $f$'s,
followed by an $a$, then followed by some string of $n-k-1$ a's or
$f$'s. A similar inequality holds for $p_1$.

We make the inductive hypothesis:
\[
|v_{f^ka}|+|v_{f^kb}| \le \delta_k, \mbox{ for k } \le n-1,
\]
where $\delta_k$ can be made as small as desired by taking $\epsilon$
small enough. We have established this for $n=1$.

Applying the triangle inequality to (\ref{eq:p0full}), we get
\[
p_0 \le \left[ |\sum_{k=0}^{k=n-2}\sum_s v_{f^kas^{n-k-1}}| +
|v_{f^{n-1}a} + v_{f^n}| \right]^2,
\]
and using the orthogonality of ${\cal S}_a$ and ${\cal S}_f$ and the
bound $|v_{f^{n-1}a} + v_{f^n}| \le 1$ we get
\[
p_0 \le \left(\sum_{k=0,s}^{k=n-2}
|v_{f^kas^{n-k-1}}|\right)^2+ 2 \sum_{k=0,s}^{k=n-2}
|v_{f^kas^{n-k-1}}| + |v_{f^{n-1}a}|^2+|v_{f^n}|^2,
\]
with a similar expression for $p_1$. Applying our inductive hypothesis
gives
\[
p_0+p_1 \le
|v_{f^{n-1}a}|^2+|v_{f^{n-1}b}|^2+2|v_{f^n}|^2+2\sum_{k=0}^{k=n-2}u_k\delta_k+\left(\sum_{k=0}^{k=n-2}u_k\delta_k\right)^2,
\]
where $u_k$ counts the number of strings $s$ for $k$. If we write this as
\[
|v_{f^{n-1}a}|^2+|v_{f^{n-1}b}|^2+2|v_{f^n}|^2 \ge p_0+p_1-y
\]
where $y$ can be made small by choosing $\epsilon$ small enough (by
the inductive hypothesis), we can subtract
\[
|v_{f^{n-1}a}|^2+|v_{f^{n-1}b}|^2+|v_{f^n}|^2 \le 1
\]
to obtain
\begin{equation}
|v_{f^n}|^2 \ge 1-\epsilon-y.
\label{eq:key}
\end{equation}
We also get
\[
|v_{f^{n-1}a}|+|v_{f^{n-1}b}| \le \sqrt{2(\epsilon+y)},
\]
and putting $\delta_{n-1}=\sqrt{2(\epsilon+y)}$ establishes the inductive
hypothesis.

Now we reintroduce measurement outcomes in the notation. Asumme $p_i$
is determined from the measurement $M_i$. We can write
\begin{equation}
p_i=|v_{f^ni}+\sum_{a,f} v_{si}|^2,
\label{eq:exact}
\end{equation}
for $i=$0 or 1. As we are assuming $p_0+p_1=2-\epsilon$ we have $p_i
\ge 1-\epsilon$ for $i=$0 or 1, and hence, taking the square root in
(\ref{eq:exact})
\[
|v_{f^ni}+\sum_{a,f} v_{si}| \ge 1-\epsilon.
\]
Using the triangle inequality and adding the results for $i=$0 and 1
gives
\[
|v_{f^n0}|+|v_{f^n1}|+|\sum_{a,f} v_{s0}|+|\sum_{b,f} v_{s1}| \ge 2-2\epsilon.
\]
As the subspaces for $M_0$ and $M_1$ are orthogonal,
$|v_{f^n0}|^2+|v_{f^n1}|^2 \le 1$, implying
$|v_{f^n0}|+|v_{f^n1}| \le \sqrt2$. So we have
\begin{equation}
|\sum_{a,f} v_{s0}|+|\sum_{b,f} v_{s1}| \ge 2-\sqrt2 -2\epsilon.
\label{eq:firsthalf}
\end{equation}
Eq. (\ref{eq:unity}) implies $\sum_{a,f} |v_{s0}|^2+\sum_{b,f}
|v_{s1}|^2+|v_{f^n}|^2 \le 1$ so from (\ref{eq:key}) we infer
\begin{equation}
\sum_{a,f} |v_{s0}|^2+\sum_{b,f} |v_{s1}|^2 \le \epsilon+y.
\label{eq:secondhalf}
\end{equation}
We now use the minimisation result that gave us theorem 3.2, namely
that, for any vectors $x_i$, $\sum_{i=1}^K |x_i|^2 \ge |\sum_{i=1}^K
x_i|^2/K$. Applying this to either of the terms on the left-hand side
of (\ref{eq:firsthalf}) and (\ref{eq:secondhalf}) gives $\epsilon+y
\ge (2-\sqrt2 -2\epsilon)^2/2^N$, and by choosing $\epsilon$ small
enough we can ensure that $N$ must be as large as we please.

\label{lastpage}

\end{document}